# Remote atomic clock synchronization via satellites and optical fibers

D. Piester[1], M. Rost[1], M. Fujieda[2], T. Feldmann[1], A. Bauch[1]

[1]Physikalisch-Technische Bundesanstalt (PTB), Bundesallee 100, 38116 Braunschweig, Germany

[2]National Institute of Information and Communications Technology, Tokyo, Japan

Correspondence to: D. Piester (dirk.piester@ptb.de)

**Abstract**

In the global network of institutions engaged with the realization of International Atomic Time (TAI), atomic clocks and time scales are compared by means of the Global Positioning System (GPS) and by employing telecommunication satellites for two-way satellite time and frequency transfer (TWSTFT). The frequencies of the state-of-the-art primary caesium fountain clocks can be compared at the level of $10^{-15}$ (relative, 1 day averaging) and time scales can be synchronized with an uncertainty of one nanosecond. Future improvements of worldwide clock comparisons will require also an improvement of the local signal distribution systems. For example, the future ACES (atomic clock ensemble in space) mission shall demonstrate remote time scale comparisons at the uncertainty level of 100 ps.

To ensure that the ACES ground instrument will be synchronized to the local time scale at PTB without a significant uncertainty contribution, we have developed a means for calibrated clock comparisons through optical fibers. An uncertainty below 50 ps over a distance of 2 km has been demonstrated on the campus of PTB. This technology is thus in general a promising candidate for synchronization of enhanced time transfer equipment with the local realizations of UTC .

Based on these experiments we estimate the uncertainty level for calibrated time transfer through optical fibers over longer distances. These findings are compared with the current status and developments of satellite based time transfer systems, with a focus on the calibration techniques for operational systems.

## 1  Introduction

Clock comparisons are one of the essential tasks of international time metrology, e.g. for the harmonization of national standards, for enabling the interoperability between satellite navigation systems, and for the dissemination of time to the public. As an internationally agreed reference the Coordinated Universal Time UTC and, more specific, the underlying International Atomic Time TAI are computed by the Bureau International des Poids et Mesures (BIPM) by using data from 391 atomic clocks distributed all over the world in 69 different institutes (as of October 2010). Most of them are National Metrology Institutes (NMIs) (Arias, 2009, Circular T).

The data involved consist of two sets to be delivered by the institutes to the BIPM. The first one is from measurements between all available clocks in each laboratory $k$ with respect to the local "physical" realization of UTC called UTC($k$). The second set is from time comparisons between the 69 UTC($k$) laboratories (see above). For these comparisons mainly two techniques are employed: In the first instance, signals from the global navigation satellite systems (GNSS) GPS and GLONASS (and in future also the European Galileo) are received and processed in different modes for global comparisons. It is a one-way technique in which time signals propagate from the satellites in known orbits through the atmosphere until they reach dedicated receivers at fixed positions. The used comparison modes depend on the equipment available at the laboratories and are called (in the order of increasing performance): "Single Channel", "Multi Channel", "P3", and "PPP". Details about the techniques can be found in the respective literature, e.g. see the papers from Arias (2007) and Petit and Ziang (2008) and references therein. All modes offer the possibility of true global comparisons, which means that clocks located at nearly every location on Earth can be compared with each other. The second technique in use is two-way time and frequency transfer (TWSTFT) (Kirchner, 1999). It has the advantage that modulated signals are exchanged between two sites, and thus propagate simultaneously through the same medium along the same path, just in opposite direction. This leads to a cancellation of most effects which have an impact on the signal delay, in particular those caused by the propagation through the troposphere and ionosphere. TWSTFT usually employs a geostationary telecommunication satellite as a relay for the signals in space. This limits the choice of locations for ground terminal installations and the operational distance to roughly 10000 km. An advantage of TWSTFT is that results of clock comparisons are available instantly after a link between two ground sites has been established. No post processing is required as compared to GNSS-based time and frequency transfer.

With these currently operational time and frequency transfer techniques, the frequencies of state-of-the-art primary caesium fountain clocks can be compared at the level of $10^{-15}$ and time scales can be synchronized with an uncertainty of one nanosecond. At present experiments for satellite based time and frequency transfer are performed or designed with the aim to reduce the so far achieved uncertainty significantly. E.g. in 2014 a new system for time and frequency transfer will be available: The Atomic Clock Ensemble in Space (ACES) will be installed at the International Space Station (ISS) revolving as time reference about every 90 minutes around the Earth. Time transfer with an uncertainty of as low as 100 ps will be possible by using the integrated ACES-Microwave-Link (MWL) (Cacciapuoti and Salomon 2009). Compared to currently operational TWSTFT the ACES MWL uses a broader bandwidth for time transfer signals. Thus also on ground dedicated technologies are needed to supply these new experiments with ultra stable reference frequency and time. Recently optical fiber based techniques show very promising results in both frequency and time transfer. Distances from campus solutions (1 km) to about 500 km have been investigated and demonstrated. Distances up to 1500 km are discussed (Piester and Schnatz, 2009). In this paper the possibility of a combination of these techniques is discussed for the application of global time and frequency transfer.

## 2  Comparison of Time Scales and Atomic Clocks

Facilities for satellite based time and frequency comparisons are standard equipment in every time laboratory. They are operated in a continuous mode and allow the determination of frequency differences between atomic clocks and phase (time) differences between time scales maintained at these laboratories. The latter requires a dedicated calibration of propagation delay in the equipment involved. Before measuring time differences via optical fibers the involved equipment has to be calibrated as well. In this section we briefly describe the two standard satellite based techniques for time transfer and finally discuss the achieved results of optical fiber time and frequency transfer based on recently performed experiments.

### 2.1  GPS Time and Frequency Transfer

GPS signal reception has become the standard tool for time and frequency transfer between time laboratories (see e.g. Piester and Schnatz, 2009). The GPS maintains a minimum of 24 satellites, in a way that at nearly each location on Earth more than 4 satellites are simultaneously visible. On each satellite an atomic clock serves as the onboard reference, which is related to the system time reference. From this reference two carriers are generated and transmitted. These signals are phase modulated by two characteristic pseudorandom noise codes, uniquely associated with each space vehicle, the so called coarse acquisition and the precise codes. As the positions of the satellites are precisely known and reported inter alia in the data stream from the satellites, ground-based clocks can be compared to GPS time with a typical precision of 10 ns or 1 ns using the coarse or precise code, respectively. Exchange of the collected data between two stations via file transfer enables the a posteriori calculation of the time differences between the two local clocks involved. If the carrier itself is used, the precision is in the range of some 10 ps (Gotoh et al., 2003), but typically subject to unknown absolute offsets.

The precision of operational GPS time and frequency transfer has been improved significantly during the last years. It is characterized by the statistical uncertainty $u_A$ for the techniques "single channel" (4.5 ns), "multi channel" (2.5 ns), "P3" (1.0 ns), and "PPP" (0.3 ns) as entitled by BIPM and is care-fully examined by the BIPM and documented in the monthly published Circular T, see e.g. issue October 2010. The accuracy of GPS *time* transfer strongly depends on the calibration procedure. The long-term instability of signal delays in the equipment making up the link determines the necessity of recalibrations, and on the other hand the accuracy of long-term frequency comparisons. For many links an systematic uncertainty $u_B$ of 5 ns has been stated, reflecting the long-lasting conservative practice. This has been put in question when Esteban et al. (2010) estimated a significantly reduced uncertainty of calibrations using a traveling GPS receiver. Recently this has been refined and uncertainties below 1 ns have been reported by Feldmann et al. (2011). There is still a need for confirmation of such small values, but an uncertainty of 1.6 ns should be feasible. It is clear that these small values for $u_B$ can be maintained over long times only when periodic re-calibrations are made.

## 2.2 Two-Way Satellite Time and Frequency Transfer

TWSTFT is the second satellite based technique which is regularly used for generating TAI. The main advantage of the two-way technique is that unknown delays along the signal path cancel out to first order because of the path reciprocity of the transmitted signals (Kirchner 1999). At present TWSTFT makes use of established communication satellite services mainly in the Ku-band (10.7 GHz to 14.5 GHz). Time signals are transmitted by relating the phase of a pseudorandom noise modulation on the carrier signal to the one-pulse-per-second input from a clock. A dedicated code is allocated to each transmitting station. The receive equipment correlates the received signal with a local replica of the signal expected from the transmitting site, and determines the time of arrival of the received signal with respect to the local clock.

With currently operational satellite link budget parameters, frequency stabilities following a $\sigma_y(\tau) = 10^{-9} (\tau/s)^{-1}$ law can be achieved, which enables frequency comparisons at the level of $10^{-15}$ at reasonable averaging times (Bauch et al., 2006). The precision at an averaging time of 1 s is about 0.5 ns. TWSTFT time comparisons can be performed with an uncertainty of about 1 ns if the internal delays in the ground station equipment is calibrated. As an example, for one time link this is usually achieved by circulating a portable reference TWSTFT station to determine the overall relative delay difference between the two local ground stations' transmit and receive equipment. Such calibrations have been done repeatedly between European stations and between PTB and the U.S. Naval Observatory (Piester et al., 2008). Breakiron et al. (2005) report for a series of intra U.S. calibration campaigns overall uncertainties down to 0.38 ns. On the other hand also values of slightly above 1 ns have been reported for recent European campaigns (Bauch et al., 2009) which lead us to conclude that using currently available equipment link uncertainties between 0.4 ns and 1.2 ns are achievable.

## 2.3 Optical Fiber Time and Frequency Transfer

The most recent developments in optical fiber time and frequency transfer have demonstrated that a further and significant reduction of uncertainty is possible. Transfer of frequency signals on fiber lengths up to more than 100 km have been demonstrated (see e.g. Grosche et al., 2009). The introduction of Brillouin amplification for the use in frequency transfer has the potential to bridge more than 250 km without intermediate amplifier stations (Terra et al., 2010). Also very promising results have been achieved for frequency transfer using standard signals for a broader application. We give two examples: a 10 MHz signal as used widely in timing laboratories (Ebenhag et al., 2008) and a 1.5 GHz signal for radio astronomy applications (McCool et al., 2008). Both approaches demonstrate frequency distributions with uncertainties below $10^{-15}$.

To cancel long term fiber length variations in time and frequency transfer the two-way method for exchanging optical signals has been investigated (Amemiya et al., 2006). Such an approach was initially proposed in the framework of network synchronization (Kihara and Imaoka, 1995).

Currently, analysis of time transfer through optical fibers (TTTOF) is investigated by different groups. Stability analyses prove the measurement precision to be at a level of 10 ps (Czubla et al. 2006) or below (Śliwczyński et al., 2010). Smotlacha et al. (2010) demonstrated their two-way system using a 740 km link. In our work we have focused on the capability of calibrated time transfer (Piester et. al., 2009b) using equipment similar to the TWSTFT scheme described before. In the two latter systems code division multiple access (CDMA) signals are employed and precisions below 100 ps and 10 ps, respectively, have been reported. The capability for calibrated time transfer on the level of 40 ps has been demonstrated in our study. A further improvement and the extension to longer distances is discussed in the next section.

## 3 Calibration and Uncertainty Evaluation of an Optical Fiber Time Transfer Link

For the operation of a time transfer link in the strict sense a suitable calibration of the propagation delays is obligatory. Here we describe a procedure which can be executed at a single laboratory before the equipment will be installed at the locations where the systems will be operated. The measurement setup is depicted in Fig. 1. We use in our setup modems which are also used for TWSTFT. Generally, we want to compare two time scales TA(1) and TA(2) generated at different distant laboratories. The modems' time-of-arrival measurements TW($i$) are given by

$$TW(1) = TA(1) - TA(2) + TX(2) + SP(2) + RX(1), \text{ and} \quad (1)$$

$$TW(2) = TA(2) - TA(1) + TX(1) + SP(1) + RX(2). \quad (2)$$

TX($i$) and RX($i$) represent the complete internal transmission and receive delay of setup $i$. SP(1) is the transmission path delay from the local setup (1) to the remote site (2). SP(2) is the propagation delay in the opposite direction through the same fiber. For a single optical fiber connection between two sites we assume complete reciprocity of this part of signal path setting SP(1) = SP(2), and thus

$$TA(1) - TA(2) = \tfrac{1}{2}[TW(1) - TW(2)] + \tfrac{1}{2}[DLD(1) - DLD(2)], \quad (3)$$

where DLD($i$) = TX($i$) – RX($i$) is the delay difference between the TX path and the RX path of one setup. For accurate time transfer we want to determine the last term in equation (3) as the calibration result. For this calibration we use a common clock configuration and adjust TA(1) = TA(2), which leads to

$$0 = \tfrac{1}{2}[TW(1) - TW(2)] + \tfrac{1}{2}[DLD(1) - DLD(2)]. \quad (4)$$

The first term is the so called common clock difference CCD(1,2) = ½ [TW(1) – TW(2)]. It is measured to determine the delay difference between both systems corresponding to a calibration value defined as

$$CALR(1,2) = \tfrac{1}{2}[DLD(1) - DLD(2)] = -CCD(1,2). \quad (5)$$

For the calibrated time transfer link between the two setups, finally installed at remote sites, we get

$$TA(1) - TA(2) = \tfrac{1}{2}[TW(1) - TW(2)] + CALR(1,2). \quad (6)$$

As noted before, a prerequisite for accurate comparisons of remote time scales is the possibility for delay calibration of the whole system. Because the optical fiber length in the final setup is unknown, a calibration test is needed to ensure the independence of the setup from the length of the used fiber which connects both laboratories. For this purpose we connected the modems to reference frequency and 1pps (1-pulse-per-second) from signal distribution equipment (FDAs) and pulse generators (DIVs) as illustrated in Figure 1. The 1pps cable connectors represent generally the time scales TA(1) and TA(2). Both TTTOF setups were then connected by two fibers which were subsequently exchanged: a 2 m short indoor fiber (Fig. 1 a)) and a 2 km long fiber (Fig. 1 b)) buried on the PTB campus. The attenuation of the two optical fibers was adjusted to be at the same level by inserting a variable optical attenuator into the bidirectional fiber path SP($i$), in order to minimize the impact of receive power dependent delay variations in the modems. The optical power was kept constant within ±0.1 dB.

The results of measurements CCD(1,2) = ½ [TW(1) – TW(2)] when switching over between long and short fibers is depicted in Fig. 2 (see Rost et al. (2010) for details of the experimental setup). The sequence comprises eight switches between the long and the short fiber. The error bars in Fig. 2 represent the standard deviation of single secondly recorded measurements. The standard deviation of the single CCD(1,2) values around the mean is only 6 ps. However, two effects were observed which have to be investigated in more detail in future: Variations at the beginning of the sequence might be a result of temperature variations and the higher standard deviation of the measurements with the short fiber may be due to instabilities caused by interference of the optical signals partly reflected at fiber connectors. A reduction of interference could be achieved if different optical wavelengths are used in the local and remote setup. Nevertheless, the small variations of below 40 ps (including error bars) obtained under different experimental conditions are promising results to comply with the aim of enabling time transfer with an uncertainty well below 100 ps to supply next generation satellite based time and frequency transfer techniques.

The operational distance between the two connected sites is limited to less than 100 km if no amplifiers are used (Amemiya et al., 2010). If longer distances should be bridged, the bidirectional signal path requires also bidirectional amplifiers. Because their internal delay is generally different for each direction a calibration of the amplifiers is also necessary. For a link with $n$ amplifiers $n$ additional calibrations have to be made. Equation (3) is extended as follows

$$TA(1) - TA(2) = \tfrac{1}{2}[TW(1) - TW(2)] + \tfrac{1}{2}[DLD(1) - DLD(2)]$$
$$+ \tfrac{1}{2}[dBA1(1,2) + dBA2(1,2) + ... + dBAn(1,2)] \quad (7)$$

where $dBAn(1,2)$ is the delay difference of the $n$-th bidirectional amplifier $BAn(1) - BAn(2)$ (see Fig. 1 c)). The values for $dBAn(i)$ can be determined as follows. The necessary amplifiers for a link are inserted into the calibration setup (common clock $TA(1) = TA(2)$). From equation (7) we get

$$0 = \tfrac{1}{2}[TW(1) - TW(2)] + \tfrac{1}{2}[DLD(1) - DLD(2)]$$
$$+ \tfrac{1}{2}[dBA1(1,2) + dBA2(1,2) + ... + dBAn(1,2)] \quad (8)$$

After changing the direction of the first amplifier the measurement is repeated:

$$0 = \tfrac{1}{2}[TW(1) - TW(2)] + \tfrac{1}{2}[DLD(1) - DLD(2)]$$
$$+ \tfrac{1}{2}[-dBA1(1,2) + dBA2(1,2) + ... + dBAn(1,2)] \quad (9)$$

Then the direction of each of all amplifiers is subsequently changed:

$$0 = \tfrac{1}{2}[TW(1) - TW(2)] + \tfrac{1}{2}[DLD(1) - DLD(2)]$$
$$+ \tfrac{1}{2}[-dBA1(1,2) - dBA2(1,2) - ... - dBAn(1,2)] \quad (10)$$

We get $DLD(1) - DLD(2)$ from adding eq. (8) and (10). We get the differential delay e.g. for the first amplifier by subtracting eq. (8) – (9). If we only want to calibrate the total link as a whole, only measurements (8) and (10) need to be combined. Thus only the uncertainties of two measurements will contribute to the overall uncertainty

$$U = \sqrt{2} \cdot u, \quad (11)$$

if we assume the same $u$ for both measurements. However, this will not allow the exchange of a single device without loosing the link calibration information. For the calibration of a whole link including $n$ bidirectional amplifiers one will need $n + 1$ common clock difference measurements. The overall uncertainty $U$ is in this case

$$U = \sqrt{n+1} \cdot u. \quad (12)$$

If we consider about $u$ = 40 ps for each delay difference determination, we expect for a 900 km link with 8 bidirectional erbium doped fiber amplifiers an uncertainty of $U$ = 60 ps following eq. (11) and $U$ = 120 ps after eq. (12).

## 4 Summary and Outlook

At present three methods exist (or are experimentally evaluated) to synchronize remote atomic clocks and time scales on the nanosecond level and below. The results for these time synchronization methods are summarized in Table 1. If one needs a time synchronization at the level of 5 ns or slightly be-low, GPS is the choice. GPS receivers are rather cheap compared to the other techniques presented, and their operational performance is almost site independent. Using the so called "all-in-view" data computation approach one can compare two sites regardingless where they are located on Earth. Most laboratories maintaining a local realization of UTC use GPS links for their connection to the international network of timing laboratories.

TWSTFT offers a lower calibration uncertainty together with a link stability both at the level of one nanosecond. Beside the superior performance some drawbacks have to be taken into account: the operational distance is limited to about 10 000 km and a geostationary satellite has to be available. It must be equipped with transponders providing the required connectivity in the visibility range of both participating stations. So the establishment of a network is more demanding. Also the number of simultaneous clock comparisons in a network is limited by the hardware used. TWSTFT ground stations are elaborate and expensive, and furthermore transponder bandwidth and time need to be purchased from the satellite operating agency, resulting in substantial running costs.

Optical fiber connections offer the best performance characteristics due to the wider bandwidth of the transmission, when a dedicated dark fiber is available. Limiting factors of time transfer via optical fibers are the repeaters needed, when exceeding 100 km distance. In the near future the very high costs for long distance fiber connections will hamper applications on the continental scale. On the other hand, campus solutions will surely be established supporting deep-space network and antenna arrays in astronomy (Calhoun et al., 2007) or in accelerators for particle physics (Kärtner et al., 2010). Connections between cities surely will be established at least for experimental purposes, e.g. to compare optical frequency standards. Broad applications requiring calibrated time synchronization at an uncertainty level well below 100 ps have not been identified so far. Nevertheless, future space experiments will require dedicated ground infrastructure for time synchronization. As mentioned above, in 2014 the ACES mission will offer the possibility to perform two-way time and frequency transfer at enhanced precision. Compared to operational TWSTFT the ACES microwave link (MWL) will allocate a broader bandwidth and therefore improve the precision to 4 ps at an averaging time of 1 s. It will be a demonstrator for enhanced TWSTFT capabilities reducing the accuracy for time transfer to 100 ps on a global scale.

A future scenario could be a combination of the presented techniques and might be used for global time comparisons: On shorter distances of several hundred kilometers laboratories might be connected via optical fibers, while these clusters are connected by intercontinental TWSTFT links and enhanced GNSS time links. The capability for TWSTFT calibrations on the global scale will be tested by circulating reference GPS receivers and mobile TWSTFT stations between the United States Naval Observatory (Washington DC), the National Institute for Information and Communications Technolgy (Tokyo) and the Physikalisch-Technische Bundesanstalt (Braunschweig, Germany). An example is depicted in Fig. 3. First steps have been done during some hours in 2010 by performing a first TWSTFT test for time comparisons around the world with the inclusion of Telecommunication Laboratories of Taiwan (Tseng, 2010). An operational global link will offer the possibility to test the accuracy of a round trip calibration to better than 4 ns, if the two GPS and the two TWSTFT calibrations' results will have uncertainties of 2 ns and 1 ns, respectively.

**References**

Amemiya, M., Imae, M., Fujii, Y., Suzuyama, S., Ohshima, S., Aoyagi, S., Takigawa, Y. and Kihara, M.: Time and Frequency Transfer and Dissemination Methods Using Optical Fiber Network, IEEJ Trans. FM, 126, 458-463, 2006.

Amemiya, M., Imae, M., Fujii, Y., Suzuyama, T., Hong, F.-L. and Takamoto, M.: Precise Frequency Comparison System Using Bidirectional Optical Amplifiers, IEEE Trans. Instr. Meas., 59, 631-640, 2010.

Arias, E. F. and Panfillo G.: International time scales at the BIPM: impact and applications, Proc. 14[th] International Metrology Congress, Paris, 2009.

Arias, E. F.: Time scales and relativity, Proc. International School of Physics "Enrico Fermi" Course CLXVI Metrology and Fundamental Constants, eds. T. W. Hänsch et al., IOS Press, Amsterdam, 367-392, 2007.

Bauch, A., Achkar, J., Bize, S., Calonico, D., Dach, R., Hlaváč, R., Lorini, L., Parker, T., Petit, G., Piester, D., Szymaniec, K. and Uhrich, P.: Comparison between frequency standards in Europe and the USA at the $10^{-15}$ uncertainty level, Metrologia, 43, 109-120, 2006.

Bauch, A., Piester, D., Blanzano, B., Koudelka, O., Kroon, K., Dierikx, E., Whibberley, P., Achkar, J., Rovera, D., Lorini, L., Cordara, F. and Schlunegger, C.: Results of the 2008 TWSTFT Calibration of Seven European Stations, Proc. European Frequency and Time Forum – IEEE Frequency Control Symposium Joint Conference, Besançon, France, 1209-1215, 2009.

Breakiron, L. A., Smith, A. L., Fonville, B. C., Powers, E. and Matsakis, D. N.: The Accuracy of Two-Way Satellite Time Transfer Calibrations; Proc. 36[th] Annual Precise Time and Time Interval (PTTI) Systems and Applications Meeting, Reston, VA, USA, 139-148, 2005.

Cacciapuoti, L. and Salomon, C.: Space clocks and fundamental tests: The ACES experiment, Eur. Phys. J. Special Topics, 172, 57-68, 2009.

Calhoun, M., Huang, S. and Tjoelker, R. L.: Stable Photonic Links for Frequency and Time Transfer in the Deep-Space Network and Antenna Arrays, Proc. IEEE, 95, 1931-1946, 2007.

Circular T: monthly publication of the BIPM, URL: http://www.bipm.org/jsp/en/TimeFtp.jsp.

Czubla, A., Konopka, J., Górnik, M., Adamowicz, W., Struś, J., Pawszak, T., Romsicki, J., Lipiński, M., Krehlik, P., Śliwczyński, L. and Wolczko, A.: Comparison of Precise Time Transfer with Usage of Multi-Channel GPS CV Receivers and Optical Fibers over Distances of About 3 Kilometers, Proc. 38[th] Annual Precise Time and Time Interval (PTTI) Meeting, 7-9, Reston, Virginia, USA, 337-345, 2006.

Ebenhag, S.-C., Hedekvist, P. O., Rieck, C., Skoogh, H., Jarlemark, P. and Jaldehag, K., Evaluation of Output Phase Stability in an Fiber-Optic Two-Way Frequency Distribution System, Proc. 40[th] Annual Precise Time and Time Interval (PTTI) Meeting, Reston, VA, USA, 117-124, 2008.

Esteban, H., Palacio, F., Galindo, F. J., Feldmann, T., Bauch, A. and Piester, D.: Improved GPS-Based Time Link Calibration Involving ROA and PTB, IEEE Trans. UFFC, 57, 714-720, 2010.

Feldmann, T., Bauch, A., Piester, D., Rost, M., Goldberg, E., Mitchell, S. and Fonville, B.: Advanced GPS Based Time Link Calibration with PTB's New Calibration Station; to be published in Proc. 42[th] Annual Precise Time and Time Interval (PTTI) Systems and Applications Meeting, Reston, Virginia, USA, 2010.

Gotoh., T., Kaneko, A., Shibuya, Y. and. Imae, M.: GPS Common View; Journal of the NICT, 50, 113-123, 2003.

Grosche, G., Terra, O., Predehl, K., Holzwarth, R., Lipphardt, B., Vogt, F., Sterr, U. and Schnatz, H.: Optical frequency transfer via 146 km fiber link with $10^{-19}$ relative accuracy, Opt. Let., 34, 2270-2272, 2009.

Kärtner, F. X., Kim, J., Cox, J., Chen, J. and Nejadmalayeri, A. H.: Femtosecond Precision Timing Distribution for Accelerators and Light Sources, Proc. IEEE Int. Frequency Control Symposium, Newport Beach, CA, USA, 564-568, 2010.

Kihara, M. and Imaoka, A.: SDH-Based Time and Frequency Transfer System, Proc. 9th European Frequency and Time Forum, Besançon, France, 317-322, 1999.

Kirchner, D.: Two-Way Satellite Time and Frequency Transfer (TWSTFT): Principle, Implementation, and Current Performance, Review of Radio Sciences 1996-1999, Oxford University Press, 27-44, 1999.

McCool, R., Bentley, M., Garrington, S., Spencer, R., Davis, R. and Anderson, B.: Phase Transfer for Radio Astronomy Interferometers, over Installed Fiber Networks, Using a Round-Trip Correction System, Proc. 40th Annual Precise Time and Time Interval (PTTI) Meeting, Reston, VA, USA, 107-116, 2008.

Petit, G. and Jiang, Z.: Precise Point Positioning for TAI Computation, Int. J. Nav. Obs., 562878, 2008.

Piester, D., Bauch, A., Breakiron, L., Matsakis, D., Blanzano, B. and Koudelka, O.: Time transfer with nanosecond accuracy for the realization of International Atomic Time; Metrologia, 45, 185-198, 2008.

Piester, D. and Schnatz, H.: Novel Techniques for Remote Time and Frequency Comparisons, PTB-Mitteilungen Special Issue, 119, 33-44, 2009.

Piester, D., Fujieda, M., Rost, M. and Bauch, A.: Time Transfer Through Optical Fibers (TTTOF): First Results of Calibrated Clock Comparisons, Proc. 41th Annual Precise Time and Time Interval (PTTI) Meeting, Santa Ana Pueblo, NM, USA, 2009b.

Rost, M., Fujieda, M. and Piester, D.: Time Transfer Through Optical Fibers (TTTOF): Progress on Calibrated Clock Comparisons, Proc. 24th European Frequency and Time Forum, Noordwijk, The Netherlands, 2010.

Śliwczyński, Ł., Krehlik, P. and Lipiński, M.: Optical fibers in time and frequency transfer Meas. Sci. Techn., 21, 075302, 2010.

Smotlacha, V., Kuna, A. and Mache, M.: Time Transfer Using Fiber Links, Proc. 24th European Frequency and Time Forum, Noordwijk, The Netherlands, 2010.

Terra, O., Grosche, G. and Schnatz, H.: Brillouin amplification in phase coherent transfer of optical frequencies over 480 km fiber, Opt. Exp., 18, 16102-16108, 2010.

Tseng, W.-H., private communication, 2010.

Table1: Achievable calibration uncertainties of time transfer links. The stated uncertainties do not comprise the long-term stability of the calibrated links.

|  | Distance (km) | Uncertainty (ns) | References |
| --- | --- | --- | --- |
| GPS | ~20 000 | < 1.6 to 5 | Esteban et al. 2010, Feldmann et al., 2011, Circular T 2010 |
| TWSTFT | ~10 000 | ~0.4 to ~1.2 | Breakiron et al., 2005, Piester et al., 2008, Bauch et al. 2009 |
| Optical Fibers | ~1500 | < 0.1 to 0.2 | Rost et al. 2010, this work |

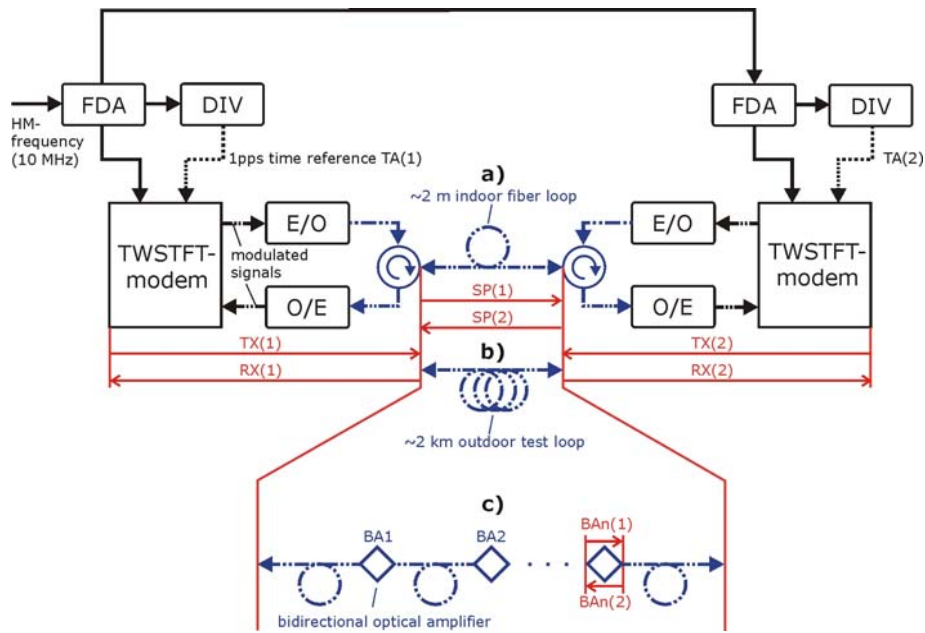

Figure 1: Calibration setup for testing the independence of the time transfer results from the length of the optical fiber. The two fibers used are depicted as a) a short fiber and b) a long fiber. On long distance connections c) a number of bidirectional amplifiers is inserted in the signal path.

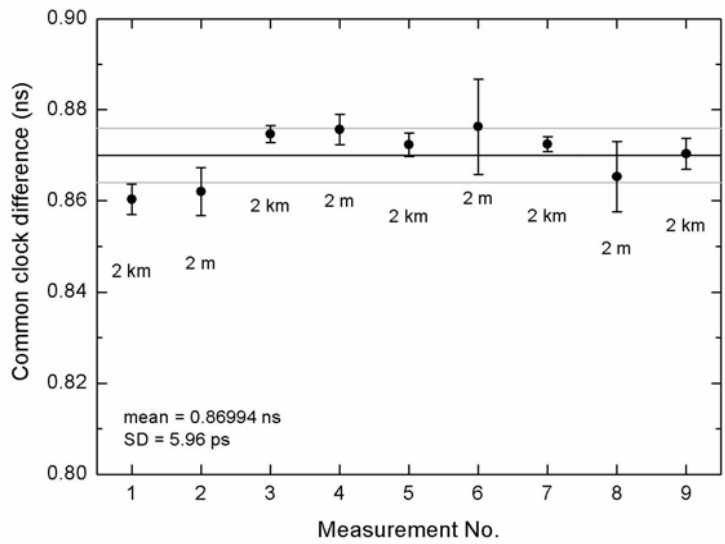

Figure 2: Display of the sequence of measurements with short fiber loop and long outdoor fiber loop (see Rost et al. 2010). The measurements are labeled with the used fiber length.

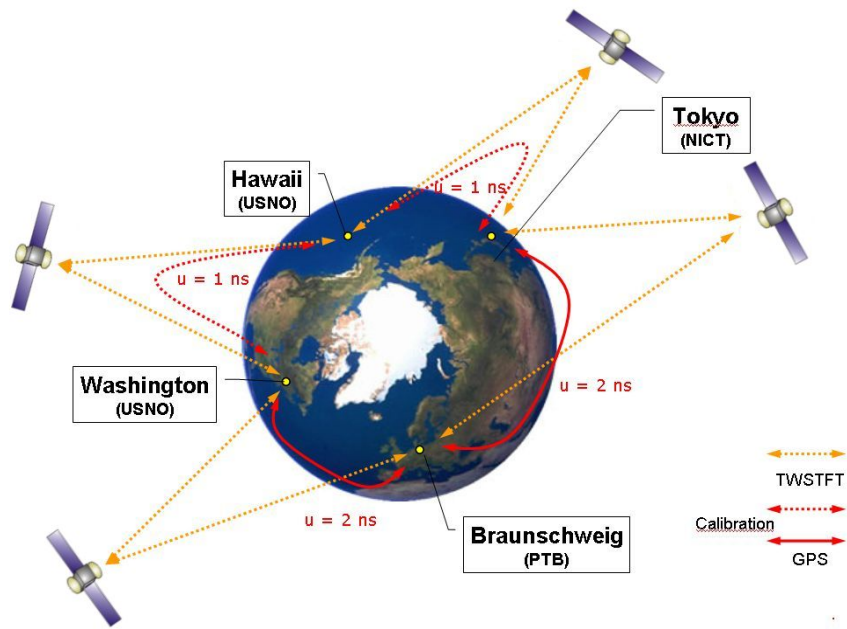

Figure 3: An example for TWSTFT around the Earth and prospective calibration uncertainties.